\documentclass[a4paper,conference]{IEEEtran}  
\IEEEoverridecommandlockouts
\usepackage{cite}
\usepackage{amsmath,amssymb,amsfonts}
\usepackage{algorithmic}
\usepackage{graphicx}
\usepackage{textcomp}
\usepackage{xcolor}
\usepackage{hyperref}
\usepackage{flushend}
\usepackage{dblfloatfix} 
\usepackage{caption} 
\def\BibTeX{{\rm B\kern-.05em{\sc i\kern-.025em b}\kern-.08em
    T\kern-.1667em\lower.7ex\hbox{E}\kern-.125emX}}

\usepackage{listings,multicol}
\usepackage{xcolor}
\definecolor{background}{HTML}{EFEFEF}
\colorlet{punct}{red!200!black}
\definecolor{delim}{RGB}{20,105,226}
\colorlet{numb}{magenta!60!black}
\definecolor{keycol}{RGB}{217,83,25}   
\definecolor{valcol}{RGB}{0,0,0}   

\lstdefinelanguage{json}{
  basicstyle=\small\ttfamily,
  numbers=left,
  numberstyle=\scriptsize,
  stepnumber=1,
  numbersep=8pt,
  showstringspaces=false,
  breaklines=true,
  frame=lines,
  backgroundcolor=\color{background},
  morestring=[b]",
  stringstyle=\color{valcol},
  moredelim=[s][\color{valcol}]{\"}{\"},  
  literate=%
   *{0}{{{\color{numb}0}}}{1}
    {1}{{{\color{numb}1}}}{1}
    {2}{{{\color{numb}2}}}{1}
    {3}{{{\color{numb}3}}}{1}
    {4}{{{\color{numb}4}}}{1}
    {5}{{{\color{numb}5}}}{1}
    {6}{{{\color{numb}6}}}{1}
    {7}{{{\color{numb}7}}}{1}
    {8}{{{\color{numb}8}}}{1}
    {9}{{{\color{numb}9}}}{1}
    {:}{{{\color{punct}{:}}}}{1}
    {,}{{{\color{punct}{,}}}}{1}
    {\{}{{{\color{delim}{\{}}}}{1}
    {\}}{{{\color{delim}{\}}}}}{1}
    {[}{{{\color{delim}{[}}}}{1}
    {]}{{{\color{delim}{]}}}}{1}
}

\usepackage{caption} 

\makeatletter
\def\lst@makecaption{%
  \def\@captype{table}%
  \@makecaption
}
\makeatother

\begin{document}

\title{Specification and Evaluation of Multi-Agent LLM Systems - Prototype and Cybersecurity Applications}

\author{\IEEEauthorblockN{Felix Härer}
\IEEEauthorblockA{
\textit{University of Applied Sciences Northwestern Switzerland}\\
Basel, Switzerland \\
felix.haerer@fhnw.ch}

}

\maketitle
\begin{abstract}

Recent advancements in LLMs indicate potential for novel applications, as evidenced by the reasoning capabilities in the latest OpenAI and DeepSeek models. To apply these models to domain‑specific applications beyond text generation, LLM‑based multi‑agent systems can be utilized to solve complex tasks, particularly by combining reasoning techniques, code generation, and software execution across multiple, potentially specialized LLMs. However, while many evaluations are performed on LLMs, reasoning techniques, and applications individually, their joint specification and combined application are not well understood. Defined specifications for multi-agent LLM systems are required to explore their potential and suitability for specific applications, allowing for systematic evaluations of LLMs, reasoning techniques, and related aspects. This paper reports the results of exploratory research on (1.) multi-agent specification by introducing an agent schema language and (2.) the execution and evaluation of the specifications through a multi-agent system architecture and prototype. The specification language, system architecture, and prototype are first presented in this work, building on an LLM system from prior research. Test cases involving cybersecurity tasks indicate the feasibility of the architecture and evaluation approach. As a result, evaluations could be demonstrated for question answering, server security, and network security tasks completed correctly by agents with LLMs from OpenAI and DeepSeek.

\end{abstract}

\begin{IEEEkeywords}
LLM, Multi-Agent System, Reasoning, Cybersecurity.
\end{IEEEkeywords}

\section{Introduction}

With the release of recent LLMs such as the DeepSeek R and OpenAI o variants, LLMs have demonstrated advancements in terms of reasoning capabilities~\cite{deepseek-ai_deepseek-r1_2025,osti_2479365}. Well-known problems and benchmarks for advanced mathematical challenges, such as AIME2024, were successfully tackled~\cite{aops_incorporated_2024_2024,brett_openai_2024}, despite being set out only recently, through novel prompting and reasoning techniques, e.g., by constructing reasoning chains of prompts and responses.

Recent advancements in models and techniques demonstrate progress on a technological level; however, their actual implementations in real-world applications are only beginning to become clear. While there does not seem to be a shortage of services and apps offering various interfaces to generative AI, the high-impact applications remain to be identified and assessed systematically. 

For instance, systematic evaluations need to determine whether specific LLMs are suitable and how they could be utilized for applications in various domains, how the LLMs and techniques compare in particular applications, in addition to advanced comparisons such as the coordination of LLM agents issuing prompt inputs and reacting to outputs for different interaction patterns, techniques, and LLMs such as established models and new reasoning models. At this point, systematic evaluations, such as benchmarks, focus on LLMs, techniques, and related aspects individually, allowing only for particular comparisons. 

In order to explore the potential for combining the specialized knowledge and capabilities of LLM agents with reasoning and prompting techniques in specific applications, the specifications in multi-agent LLM systems require further exploration to assess their potential in applications and for systematic evaluations at the application level. This paper reports on the initial results of experimental research towards evaluating multi-agent LLM systems through prototyping. The resulting artifact is an LLM execution system, initiated in 2023~\cite{haerer23_cmi}, which has been extended to specify and evaluate multi-agent LLM applications. In particular, an extended architecture and a defined specification support multi-agent systems that can combine multiple, specialized LLMs and support prompting and reasoning techniques in the execution of tasks. A system overview is presented in this paper, comprising the architecture and specification language together with test cases for cybersecurity tasks. The test cases evaluate network and server security tasks with agent specifications using commercially and openly available state-of-the-art LLMs by OpenAI and DeepSeek. 

In the remainder of this paper, Section~\ref{sec:background-related-work} introduces background and related work, the architecture and specification are discussed in Section~\ref{sec:architecture}, and these are demonstrated with cybersecurity test cases in Section~\ref{sec:cases}. Section~\ref{sec:conclusion} concludes.

\section{Background and Related Work}
\label{sec:background-related-work}

Agents can carry out actions to complete complex tasks, individually or jointly in the loosely coupled structures of multi-agent systems~\cite{maldonado_multi-agent_2024,li_survey_2019}. In principle, agents act on their own and commonly involve functional components depending on the environment, e.g., agents using sensors in industrial environments or software agents with executable functions and corresponding data. Generally, tasks are orchestrated in the agent system by interaction among agents, such as invoking specialized agents with specific functions, knowledge, or data. In the context of Large Language Models (LLMs), these models serve as the foundation for completing tasks~\cite{dong_survey_2024,guo_large_2024}, utilizing prompts and responses that invoke functions, read or write data, or perform system-specific actions to complete tasks and solve problems. 

LLMs based on Generative Pre-trained Transformers (GPT) predict sequences of tokens that are mapped to entities such as words or word fragments, pixel or related image and video feature representations, and other data entities~\cite{kumar_large_2024,RayChatGPT2023,RothmanTransformer2022}. These values are encoded and embedded to be represented in a high-dimensional vector space, in addition to the positions in the sequence. Passing through multiple layers of the GPT architecture, in the form of transformer blocks, attention and feed-forward network components predict tokens and probabilities based on attention values~\cite{VaswaniAttention2017} and sampling for the selection of any following token. Since attention relates to prior positions in the sequence with similar attention, these positions represent and find related concepts, rather than using only probabilities to predict the next token. For this reason, the semantic capabilities of agents have recently been greatly enhanced, as evidenced by recent model releases.

Recently, OpenAI and DeepSeek demonstrated reasoning models, where the network components are trained using reinforcement learning, along with prompting and reasoning techniques to produce chains of logically following reasoning steps~\cite{deepseek-ai_deepseek-r1_2025}. Reasoning models gain the ability to produce one or multiple subsequent answers, which might be arranged as a tree or graph, and to evaluate the answers in a subsequent reasoning step. By considering one or multiple answers in conjunction with the context of previous steps, this input enables the LLM to evaluate coherence and select answers to continue accordingly. In tree or graph structures, multiple paths might be explored before selecting a coherent line of reasoning~\cite{luo_graph-constrained_2024,prasad_receval_2023}. Thus, reasoning models allow for reflection and also introspection, e.g., visible in the openly available DeepSeek models that generate a "<think>" tag for reasoning that is closed before the final answer is returned. 

Reasoning chains and techniques such as Chain-of-Thought (CoT) can achieve behavior similar to reasoning and have also demonstrated recent advancements~\cite{lin_constrained_2024,wang_self-consistency_2023}. CoT is a well-known approach that issues prompts in order to generate the responses that reflect on the original question and the conversation context; thus, it is not limited to reasoning models. Several variants and further techniques exist. Notably, augmentation of prompts, e.g., using retrieval augmented generation, can add information or knowledge relevant to a prompt and improve answers~\cite{2025rag}. Formulating constraints with zero- or few-shot prompting~\cite{FillHVBRMCB24,lin_constrained_2024,luo_graph-constrained_2024} allows the LLM to learn from examples that are concrete instances or provide abstract guidance or structures. Constraints might also be placed on the content of answers, where related aspects are required to adhere to defined criteria, being less prone to hallucinations. In cases where structural or syntactical properties are constrained, the LLM tends to replicate and instantiate the structure or syntax correctly. 
\vspace{10mm}

While these techniques can be applied generally, they are not considered explicitly in LLM or agent specifications, e.g., for platforms or evaluations of LLMs and agents. Several domain-specific benchmarks exist for comparing LLMs, such as for cybersecurity applications~\cite{tihanyi_cybermetric_2024,bhusal_secure_2024,liu_cyberbench_2024} or sub-disciplines such as threat intelligence~\cite{alam_ctibench_2024}. The results generally indicate high accuracy for larger, state-of-the-art models and point out the future potential in the area. This view is also shared by two recent surveys~\cite{zhang_when_2025,sai_generative_2024} that provide a comprehensive overview. 

\vspace{2mm}

While multi-agent LLM systems are beginning to appear in practice, the specification of multi-agent LLM systems is still not well understood and is underexplored. Especially the specification by a format or language that allows for constraints, reasoning chains, and related prompting and reasoning techniques with multiple agents is not evident in the literature. Thus, the specification and platform, demonstrated in this explorative research paper, are intended to enhance the understanding of multi-agent systems and inform future LLM responses and agent development through systematic evaluations. E.g., by evaluating reasoning techniques, LLMs designed for reasoning or not designed for reasoning, and combining techniques selectively depending on capabilities and knowledge.

\section{Multi-Agent LLM System Architecture and Specification}
\label{sec:architecture}

The following subsections discuss the overall concept based on high-level requirements, a system architecture realizing these requirements, and a specification for multi-agent LLM systems.

\subsection{High-Level Requirements}
\label{subsection:concept}

In order to outline the concept of a multi-agent LLM system, the following high-level requirements are defined to establish the architecture:

\begin{figure*}[ht]
    \centering
    \vspace{-2mm}
    \includegraphics[width=0.85\linewidth]{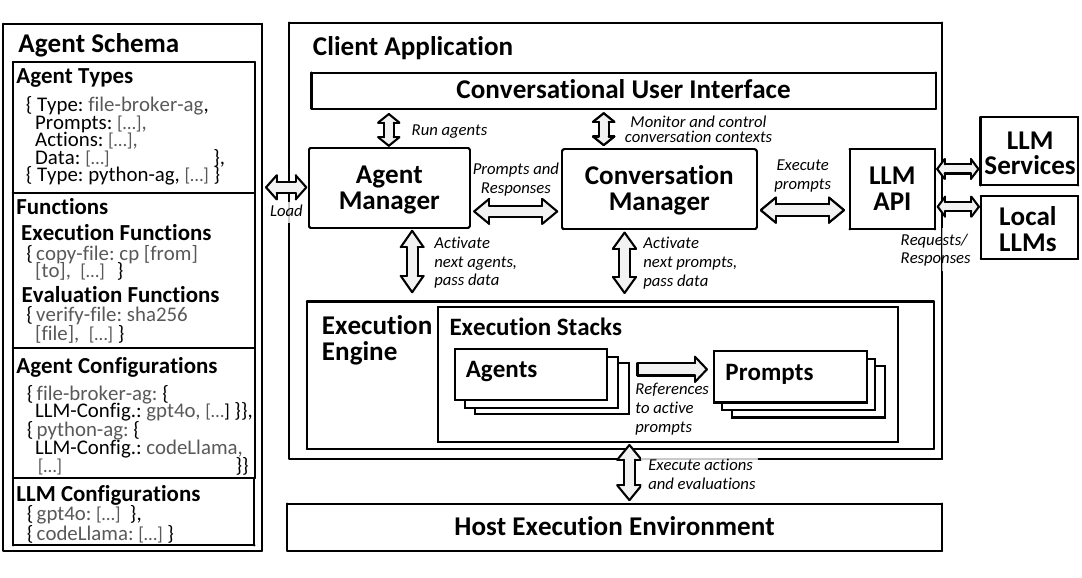}
    \caption{Architecture describing the components of the client application together with the specification in the agent schema.}
    \label{fig:architecture}
    \vspace{-5mm}
\end{figure*}
\begin{itemize}

    \item[1.] Specification of the access to open source and commercial LLMs through interfaces and parameters with an API or a local LLM runtime such as Ollama.

    \item[2.] Specification of agent types for the instantiation of individual agents, which invoke LLMs, task actions, and other agents with behavior depending on LLM outputs, action execution results, results of other agents, and the evaluation of results.

    \begin{itemize}

        \item[2.1]  Specifying LLM prompts in the conversation contexts of one or multiple agents in data structures that support prompting and reasoning techniques in terms of sequential prompts, multiple chains, and constraints. 
        
        \item[2.2]  Specifying task actions such as executable commands and functions together with their availability in each agent type. For an instantiation, initial agent data might be defined at build-time to initialize and then execute actions at run-time. The action execution inputs and outputs as well as agent data must be inserted into the conversation contexts.
        
        \item[2.3]  Evaluating results at run-time, originating from LLM responses, outputs of action executions, or other agents, i.e., the evaluation through other LLM agents. An evaluation must conclude whether a result satisfies defined criteria, e.g., computation or calculation results falling in defined classes or reaching an expected value.
      
        \item[2.4]  Invoking prompts, action execution for an agent or other agents at run-time unconditionally or conditionally based on previous results and evaluations.
        
    \end{itemize}
\end{itemize}

The requirements related to prompting and evaluation concern the multi-agent systems' capability to support specific techniques, such as chaining or constraining the reasoning in the context of action executions. For example, constraints can be placed on prompts and responses regarding their format, e.g., text strings, integer or decimal values, patterns, or complex syntax rules of formal languages or programming languages. In this case, the evaluation encompasses a syntax evaluation of the LLM responses before execution by, e.g., a Python-capable agent.

\subsection{Execution Architecture}

The following architecture extends previous research on LLM execution~\cite{haerer23_cmi} to a multi-agent system architecture with a corresponding specification, realizing the high-level requirements. Architecture and specification are shown in Figure~\ref{fig:architecture} and in the form of a client application with its sub-systems and an agent schema for specifying agent types with related functions and configurations. 

The sub-systems manage agents and the conversation through the components \textit{Agent Manager}, \textit{Conversation Manager} and \textit{Conversational UI}. Given an agent schema, \textit{Agent Manager} requests user interface (UI) actions in \textit{Conversational UI} such as agent settings and the loading of prompts in \textit{Conversation Manager}. When running an agent, \textit{Conversation Manager} activates the agent in the \textit{Execution Engine} by pushing it onto a stack for execution together with the referenced prompts, activated by the \textit{Conversation Manager}. An active prompt is executed by \textit{Conversation Manager} through \textit{LLM API} while loading prompts and streaming responses to \textit{Conversational UI} for monitoring and conversation control. The architecture and prototype support external LLM services with common APIs such as the OpenAI API and Replicate, in addition to local LLM applications such as Ollama. Active agents reference their next prompts, both placed on separate stacks. In the stack-based execution, any prompt at the top of the stack is (1.) processed by an LLM via \textit{Conversation Manager}, (2.) the generated response is received by the top-most agent, and (3.) the agent executes all specified functions on the \textit{Host Execution Environment} (HEE) and supplies the response as input. The functions referenced in a prompt are defined in the form of execution or evaluation functions, where execution functions relate to actions that are executed first, e.g., a "copy-file" function of the agent file-broker-ag, running in an Ubuntu Server HEE. After executing all actions, the specified evaluation functions, such as "verify-file", will be called and determine the result of the execution. For example, such a function might compute a file checksum using SHA-256 and match the result with an expected value, resulting in a true evaluation result in case of a match or in a false evaluation otherwise. Further examples are discussed in Section~\ref{sec:cases}.

\subsection{Agent Schema Specification}
\label{sec:spec-schemata}

\begin{lstlisting}[float=*, floatplacement=h!,language=json,firstnumber=1,caption={Agent-Type specification for question answering in a Security Q\&A Agent and for network scanning tasks in a Network Security Agent. Based on questions and answers from the CyberMetric dataset~\cite{tihanyi_cybermetric_2024}, performance is evaluated for the LLM and the configuration of the Q\&A agent. For task execution, performance is evaluated based on expected classes of results, defined by values or patterns. E.g., the Network Security Agent is expected to find open ports at host 10.11.1.24.},captionpos=b,label={lst:case-1-spec}]
[ { "id": "Security-Q&A-Agent",
    "prompts": [ { 
        "prompt": "In TCP/IP networking, which protocol is used to hold network addresses and routing information in a packet?", 
        "answers": { "A": "HTTP", "B": "IP", ... }, 
        "answer": "B" }, ... ],
    "prompt-template": "Question: [question]\n\nOptions:\nA) [answers/A]\nB) [answers/B]\nC) [answers/C], ...",
    "evaluate": { 
        "result-classes": [ { 
        "class": "A", "pattern": "ANSWER: A", "eval-expected": "correct", "eval-unexpected": "incorrect" }, ... ] } },
  { "id": "Network-Security-Agent",
    "prompts": [ { 
        "prompt": "Scan the local network [ipv4-network] for reachable hosts with commonly exposed ports. Use the nmap module.", 
        "actions": ["write-to-file", "extract-ip-scan-results"], 
            "expected-value": "10.11.1.24" }, ... ],
    "actions": ["write-to-file", "extract-code", "evaluate-syntax-shell", "execute-shell"],
    "data": { 
        "report-file": "network-report.txt", "ipv4-network": "10.1.1.0/24" }, ... } ]
\end{lstlisting}

For specifying multiple agents, executable tasks, and evaluation functionality, a schema-based approach is realized, drawing from conceptual modeling and object orientation. On an abstract schema level, the schema provides a template for agent types, their related functions, and configurations. The specification in Figure~\ref{fig:architecture} shows an overview of the domain-specific language structure with examples. In the prototype, the concrete syntax is realized in JSON format.

An \textit{Agent Schema} encompasses \textit{Agent Types}, \textit{Functions}, \textit{Agent Configurations}, and \textit{LLM Configurations}. Agents are instantiated based on \textit{Agent Types} that specify the executable prompts and actions together with data in the form of a key-value data structure. Actions specify the behavior of agent types, defining the actions an agent can carry out to complete tasks. They reference \textit{Execution Functions} that consist of executable commands with inputs and outputs denoted in data variables and \textit{Evaluation Functions} for the evaluation of results produced by an agent. At run-time, function inputs are the LLM response and the data, held in the key-value data structure in the agent on the stack. Functions read and write to the data structure, e.g., they might extract code blocks from an LLM response, evaluate the syntax of generated Python code, or execute arbitrary code. \textit{Evaluation Functions} operate on the LLM response and execution outputs. They specify conditions, values, patterns, and classes to evaluate an expected or unexpected result. The action specification on the agent level applies to all prompts and may be overwritten on the prompt level. There, it may be specified unconditionally or in a case statement conditionally, in case the denoted result class applies.

Furthermore, \textit{Agent Configurations} specify configurations required to initialize each agent with its defined type at run-time, including function parameters, data, and a reference to an LLM configuration defined in \textit{LLM configurations}, where LLM and API parameters are set. 

\section{Test Cases for Cybersecurity Applications}
\label{sec:cases}

This section discusses exemplary test cases with an evaluation for cybersecurity tasks involving an agent specification with OpenAI and DeepSeek models. The test cases encompass question answering as well as server and network security tasks that apply prompting and reasoning techniques. These test cases aim at evaluating the feasibility of the architecture and specification.

\subsection{Q\&A Specification by Templating}

Answering a set of questions with the expectation of open or pre-defined answers is a foundational application for LLMs. The LLM is presented with a series of questions, where each response is evaluated against an expected result by an evaluation function, e.g., determining a match with a value or pattern individually or in pre-defined classes. Case 1 applies this concept in a series of 10 cybersecurity questions from the CyberMetric Q\&A dataset~\cite{tihanyi_cybermetric_2024} as an example. Listing~\ref{lst:case-1-spec} shows the definition of an agent type Security Q\&A Agent that is prompted with a Q\&A specification. A template defines the format for questions and answers and specifies loading questions and answers from a file or inline. Evaluation functions are defined by classes A to D, matching the expected results in the dataset. 

According to the specification, the evaluation computes result classes by agent and prompt. As shown in Table~\ref{tab:results-cases}, results indicate all LLMs managed to complete the 10 test questions correctly, indicating basic cybersecurity knowledge and basic question answering capabilities. The source code and the complete results are published online\footnote{\href{https://github.com/fhaer/multi-agent-llm-system}{https://github.com/fhaer/multi-agent-llm-system}}.

\begin{figure*}[!t]
    \centering
    \vspace{-2mm}
    \includegraphics[trim={0 0 0 0.2cm},clip,width=1.0\linewidth]{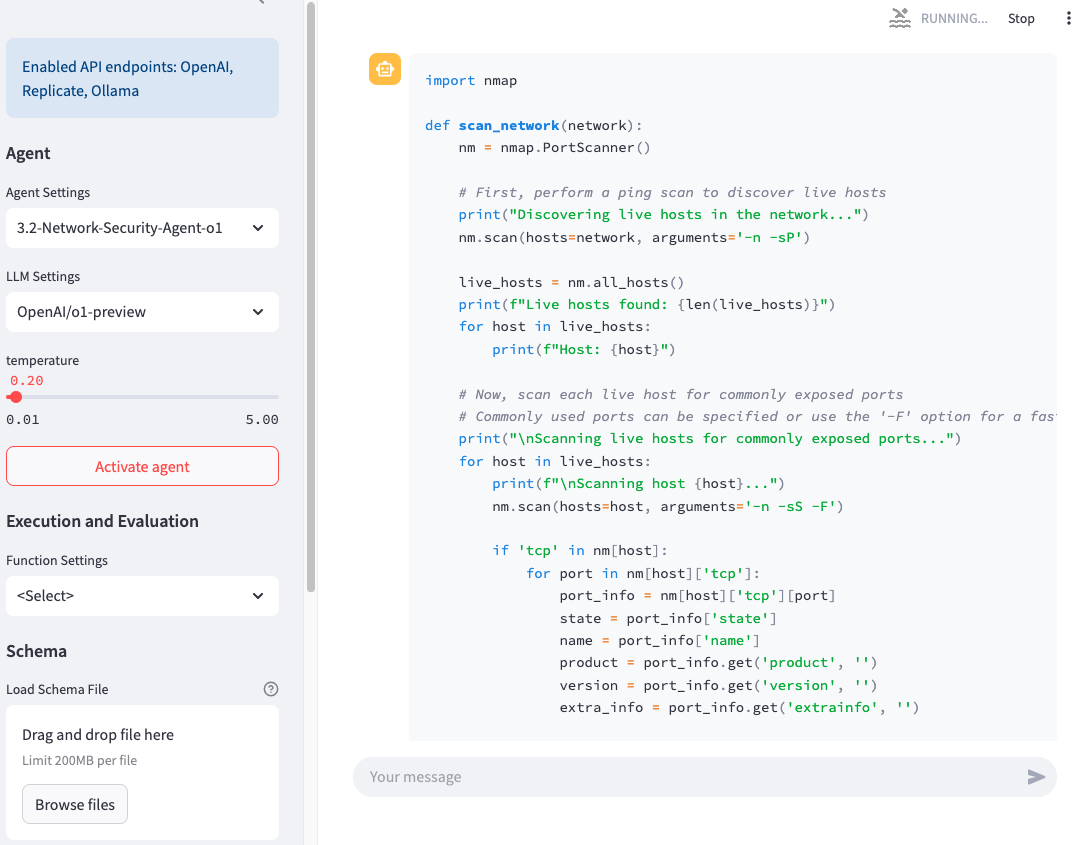}
    \caption{Excerpt of a Python script generated for the network scanning task in Listing~\ref{lst:case-1-spec} within the user interface of the client application. The code was generated and executed by a Python-capable agent of the type Network Security Agent. }
    \label{fig:case-scan}
\end{figure*}

\subsection{Task Execution Specification}

For executing tasks, their individual actions are specified for an execution environment. Case 2 sets up server and network security agents with eight tasks aimed at typical activities, including assessing firewall configurations and scanning a local network in Ubuntu Linux 24.04. Listing~\ref{lst:case-1-spec}, previously introduced, shows the type Network Security Agent with instructions to generate a network scanning script, executed according to the defined actions. Further instructions on using Python in the Ubuntu environment were given in the system prompt. The Python code generated and executed by the network security agent is shown in Figure~\ref{fig:case-scan}. 

\begin{table*}[!t]
\centering
\caption{Results of correct and incorrect task executions by the specified agents. Agent IDs denote the agent LLMs OpenAI GPT-4o (4o), OpenAI o1-preview (o1), DeepSeek R1 (R1), and DeepSeek R1 Distilled Llama 8B (R1D).}
\includegraphics[trim={0 0 0 0.2cm},clip,width=0.55\linewidth]{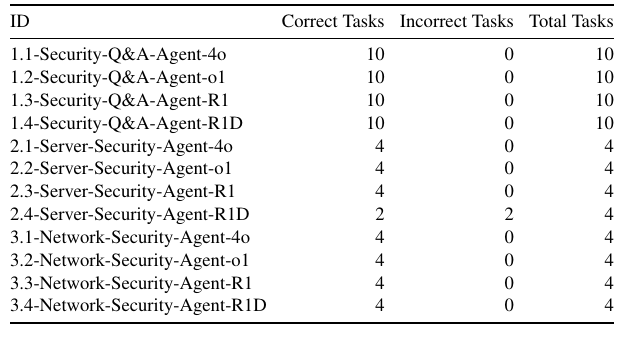}
\label{tab:results-cases}
\vspace{-0.05cm}
\end{table*}

\begin{figure*}[!b]    
  \centering
\begin{lstlisting}[language=json,firstnumber=1,caption={Agent-Type specification, where a Server Security Agent invokes an Audit Report Agent and passes data.},captionpos=b,label={lst:case-3-spec}]
[ { "id": "Server-Security-Agent",
    "prompts": [ {
       "prompt": "...", 
       "invoke": { 
           "agent-of-type": "Audit-Report-Agent", 
           "prompt-id": 1, 
           "data-keys": ["ipv4-address", "ipv4-network", "scan-result"] } }, ... ] },
  { "id": "Audit-Report-Agent",
    "prompts": [ { 
       "prompt": "Create a report of the findings for the server with IP address [ipv4-address] in the network [ipv4-network]. Consider each of the hosts found in the scan result, identify potentially vulnerable services, and give recommendations to address     potential vulnerabilities. \n\nScan result:\n\n[scan-result]", ... } ], ... } ]
\end{lstlisting}
\end{figure*}

\subsubsection{Augmentation and Constraints} In the processing of prompts and responses, augmentation and constraint techniques are used for task executions. Prompts are augmented at run-time with agent data or results of previous executions as specified by agent data keys, e.g., ipv4-network. In this way, retrieval augmented generation (RAG) is possible also for information or knowledge retrieval, e.g., by actions containing database or knowledge graph queries. Constraints reduce the solution space and narrow possible answers by specifying aspects of the response structure or content. When executing actions, the constraints specified in prompts are enforced, e.g., requesting and requiring a response in a specific data type such as an IPv4 address, a language such as Python, or data format such as JSON. 

Here, the agent executes generated Python scripts according to actions that check the syntax, execute, and further process results. Thus, the agent-type must rely on an LLM with sufficient domain knowledge, programming capabilities for Python, and the Ubuntu 24.04 environment. For the server and network security tasks, Table~\ref{tab:results-cases} shows the evaluation results. 

The tested LLMs are state-of-the-art models at the time of this writing, except for the relatively small DeepSeek-R1-Distilled LLM (R1D), which has only 8 billion parameters and was added for comparison. All of the larger state-of-the-art models completed each of the 4 network and server security tasks correctly. R1D managed to complete all 4 network security tasks and 2 of the 4 server security tasks correctly. In the first instance of incorrect task execution, R1D constructed an incorrect shell command for retrieving the system firewall configuration, resulting in a syntax error (server security task 2). In the second case, R1D used an incomplete command for creating a system report (server security task 4). The complete results are published online. 

\subsubsection{Reasoning Chains} Based on a series of prompts and responses, reasoning chains can be constructed in combination with constraints~\cite{lin_constrained_2024,luo_graph-constrained_2024}. In addition to Chain-of-Thought (CoT), which aims to issue prompts that generate subsequent reasoning steps with their responses, the chain can also be constructed explicitly by relying on pre-defined structures. Constraints ensure their existence and allow prompting the LLM of an agent with the next step. This approach is especially relevant in multi-agent scenarios, where another, specialized agent might be invoked based on data collected beforehand. 

\begin{figure*}[p]
    \centering
    \vspace{-2mm}
    \includegraphics[trim={0 0 0 1.2cm},clip,width=1.00\linewidth]{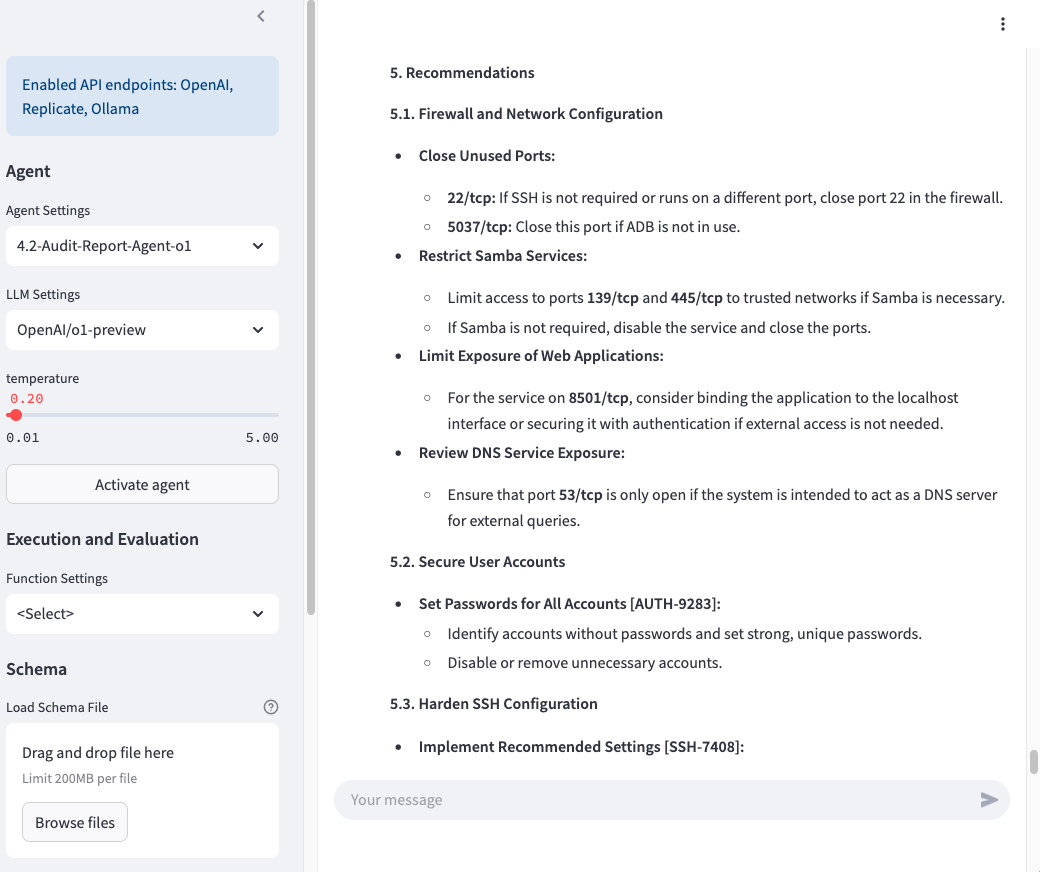}
    \caption{Excerpt of a security report generated by an Audit Report Agent after the completion of server-side tasks by a Network Security Agent in the user interface of the client application. The full results are published in the online repository at \url{https://github.com/fhaer/multi-agent-llm-system}.}
    \label{fig:case-report}
    \vspace{-0.5mm}
\end{figure*}

As an example, Listing~\ref{lst:case-3-spec} shows the Server-Security-Agent type that establishes an IP address, network, and scan result data after a series of prompts and invokes an agent of type Audit-Report-Agent. On the basis of these prompts, the agent is requested to create a report using the passed data. Further actions may impose constraints or specific conditions that influence which agent is invoked or the data passed to the agent. Figure~\ref{fig:case-report} shows an excerpt of the generated report. 

\section{Conclusion}
\label{sec:conclusion}

This paper presented the initial results of experimental research on the specification of multi-agent LLM systems. In this first system overview, the architecture and specification language were demonstrated and applied in test cases related to question answering as well as network and server security tasks. The test cases indicate the feasibility of the architecture and specification, showing the completion of tasks with agent specifications based on state-of-the-art commercially and openly available LLMs. In particular, the completion of the test cases shows the potential of LLM agents for software-based tasks in applications. Factors enabling the task completion are a combination of (1.) involving the knowledge capabilities of specialized agents such as for cybersecurity and code generation, (2.) utilizing reasoning techniques such as reasoning chains, and (3.) agents executing the inferred tasks through software actions. In future research, this combination will allow the exploration of novel multi-agent LLM applications with advanced reasoning requirements and, overall, support systematic evaluations in cybersecurity and other domains.

\clearpage

\clearpage

\bibliographystyle{IEEEtran}
\bibliography{./literature.bib}

\end{document}